\documentclass[12pt]{iopart}
\usepackage{graphicx}
\begin{document}
\title{Symmetrization, quantum images and measurement}
\vspace{4cm}
\author{Fariel Shafee}
\address{ Department of Physics\\ Princeton University\\
Princeton, NJ 08540\\ USA.}

\ead{fshafee@princeton.edu } \vspace{4cm}

\begin {abstract}
We argue that symmetrization of an incoming microstate with similar
states in a sea of microstates contained in a macroscopic detector
can produce an effective image, which does not contradict the
no-cloning theorem, and such a combinatorial set can then be used
with first passage random walk interactions suggested in an earlier
work to give the right quantum mechanical weight for measured
eigenvalues.
\end{abstract}


\maketitle

\section{Introduction}

In a previous work \cite{SH1} we have presented a picture of the
transition of a superposed quantum microstate to an eigenstate of
a measured operator through interactions with a measuring device,
which are random in the sense of the stochasticity introduced by
the large number of degrees of freedom of the macrosystem, and not
due to any intrinsic quantum indeterminism. Random walks
\cite{ZU1} have long been a favorite sports enjoyed by many
quantum physicists in search of a rationale for quantum
indeterminism \cite{BU1}. Different stochastic models for
transitions to collapsed states on measurement have been presented
by many authors \cite{PE1,GH1,DI1,GI1,AD1}. However, in our work
we made the novel departure of using first passage walks
\cite{RE1} which lead to a dimensional reduction of the path in
simplicial complexes to simplexes of lower dimensions by turn, a
possible feature also noted very recently by Omn\`{e}s \cite{OM1}.
In the work cited we appealed to heuristic arguments in analogy
with electrodynamic images. In the present work we try to justify
the emergence of image-like subsystems in a macrosystem from
quantum symmetry principles.

\section{Symmetrization and Interactions}

Interactions between systems may be due to Hamiltonians connecting
operators that explicitly connect components of different systems,
or they may be due to symmetrization or anti-symmetrization of the
states of the systems involved. For fermions, exchange interaction
yields the exclusion principle, which may have more dominant
effects than a weak potential in a many-particle system. For
bosonic systems condensation at low temperatures indicate the
creation of macro-sized quantum states. Unlike the unitary
time-dependent operators representing the explicit interactions
between systems through the Hamiltonian, (anti-)symmetrization has
no explicit time involvement, and a system includes the
(anti-)symmetrization of the component subsystems \emph{ab
initio}, which continues until the states change and lose their
indistinguishability. Alternatively, (anti-)symmetrization comes
into action as soon as an intermediate or final state is produced
involving identical particles, even when the initial system might
not have had any. The process therefore is apparently a discrete
phenomenon, going together with the abrupt action of the creation
or annihilation of particles in field theory.

In terms of first quantized quantum mechanics, we understand the
permutative (anti-)symmetry properties of identical microstates
(particles) in terms of the separation of the co-ordinates). Two
identical microsystems labeled $1$ and $2$ in a particular states
$a$ and $b$ has the combined (anti)symmetric wave function

\begin{equation}\label{eq1}
\psi(1,2)_{ab} =\frac{1}{\sqrt{2}}[\psi(1)_a  \psi(2)_b  \pm
\psi(2)_a \psi(1)_b]
\end{equation}

In practice the labels $1$ and $2$ for the two particles usually
refer to the concentration of the two particles in two different
regions of space, for example, near two attractive potential
centers. So, the labels $1$ and $2$ are actually also interpretable
as parameters for two different states, and \cite{YO1} it is
possible to combine the two sets of labels into a single set, say
$\alpha$  and $\beta$  and demand that

\begin{equation}\label{eq2}
\psi(\alpha, \beta)  = \pm \psi (\alpha', \beta')
\end{equation}

where the sign for fermionic systems depends on the number of
interchanges needed to obtain the parameter sets $\alpha'$ and
$\beta'$ from the unprimed sets and for the bosons it is of course
always positive.

Even with \emph{ab initio }symmetry built-in, it is well-known
that a state can dynamically evolve from a nearly factorized
separable product state to a fully symmetric entangled state as
the overlap becomes high from nearly zero when the two subsystems
(particles) were well-separated initially. If we know that the
incoming particles labeled $1$ and $2$ were in states $a$ and $b$
at large separations then,

\begin{eqnarray}\label{eq3}
\psi(x_1,x_2,a,b) = \frac{1}{\sqrt{2}} [\psi(x_1,a)\psi(x_2,b) \pm
\psi(x_2,a) \psi(x_1,b)]  \nonumber \\
\sim  \frac{1}{\sqrt{2}} \psi(x_1,a)\psi(x_2,b)
\end{eqnarray}

for $|x_1 - x_2|$ large, as the second term is small .

If the states $a$ and $b$ are identical, then it is well-known
that this exchange interaction for bosons gives an effective
attractive interaction for small $|x_1 - x_2|$, as we get simply
$\sqrt{2}$ times a single wave function, whereas for fermions it
becomes highly repulsive as the antisymmetry produces the
exclusion principle.

\section{State of the Detector}

We shall consider a  detector as macrosystem which consists of a
large number of microsystems identical with the microsystem to be
detected, but in all possible different states, including the
incoming state to be detected, so that initially it appears like a
neutral unbiased system with respect to the state of the incoming
microsystem. This picture is comparable to that of a sea of quarks
of all flavors and colors in a quark bag, or even the similar
content of a neutral vacuum when considering vacuum polarization
contributions. To maintain the quantum number of the vacuum, i.e.
to give a singlet with respect to all possible
symmetry/classification groups, all these states occur paired with
conjugate anti-states (group theoretically inverse elements):

\begin{equation}\label{eq4}
\Psi_D = \sum_a \psi_{Da} \bar{\psi}_{Da}
\end{equation}

where the label $D$ indicates states with positional peaks inside
the detector. The expression above is the simplest spectral
decomposition for our purpose. In general there will also be
simultaneous multiple state/anti-state pairs, which will introduce
new numerical factors from combinatorics, but will not change the
relative strengths of interactions between the incoming microstate
(S) and the pairing anti-states of the detector (D), which is the
crucial part of our measurement picture.

\section{System-Detector Symmetry}

For a bosonic microsystem system being detected, if it is in the
state $\psi_{Si}$, symmetrization with the detector states gives

\begin{equation}\label{eq5}
\psi_{SD}  = \frac{1}{\sqrt{2N}} \sum_{j\neq i} ( \psi_{Si}
\psi_{Dj} + \psi_{Sj} \psi_{Di} ) \bar{\psi}_{Dj} +
\frac{1}{\sqrt{N}} ( \psi_{Si} \psi_{Di}  \bar{\psi}_{Di})
\end{equation}

when there are $N$ states uniformly distributed in the detector,
including the state $i$. Normalization is ensured by the
orthogonality of the states, when the coefficients are as chosen.

However, if the microsystem was well-separated from the detector and
symmetrization was not invoked, the product state in a product space
would be, with the macroscopic detector still containing a
superposition of all possible states:

\begin{equation}\label{eq6}
\Psi_{SD0}  = \frac{1}{\sqrt{N}} \sum_{ all j} ( \psi_{Si} \psi_{Dj}
\bar{\psi}_{Dj})
\end{equation}

Since the functions $\psi_{Si}$ and $\psi_{Di}$ for an identical
microsystem in the same state may both actually represent the
observed incoming microsystem in Eq. \ref{eq6}, we can rewrite Eq.
\ref{eq5} as

\begin{equation} \label{eq7}
\Psi_{SD}  = \frac{1}{\sqrt{2}} (\Psi_{SD0} +\Psi_{DS0}) +
\frac{1}{\sqrt{N}}(1-\sqrt{2}) \psi_{Si} \psi_{Di} \bar{\psi}_{Di}
\end{equation}.

Here both $\Psi_{SD0}$ and $\Psi_{DS0}$ represent an incoming
particle in the state $\psi_i$ and its noninteracting product with
the detector. Hence, the extra term $\psi_{Si} \psi_{Di} \psi_{Di}$
represents the 'exchange interaction' resulting from the
entanglement of the microstate with the detector.

For incoming fermionic systems the arguments are similar, but
somewhat more complicated. In this case anti-symmetrization gives

\begin{equation} \label{eq8}
\Psi_{[SD]}  =  \frac{1}{\sqrt{2(N-1)}} \sum_{j\neq i} ( \psi_{Si}
\psi_{Dj} - \psi_{Sj} \psi_{Di} ) \bar{\psi}_{Dj}
\end{equation}

Since the detector includes all other states but must exclude the
state $\psi_i$ due to anti-symmetrization (exclusion principle), we
can actually consider the sums in Eq. \ref{eq8} as involving
hole-antihole pair states $\psi^h_{Di}\bar{\psi^h}_{Di}$
corresponding to $\psi_i$.  So we have for the combined system of
the incoming microsystem $\psi_i$ and the detector:

\begin{equation} \label{eq9}
\Psi_{[SD]}  \sim (\psi_{Si} \psi^h_{Di} - \psi_{Di}\psi^h_{Si})
\bar{\psi^h}_{Di}
\end{equation},

with the definitions:

\begin{eqnarray} \label{eqn10}
\psi^h_{Di}\bar{\psi^h}_{Di}= \sum_{j\neq i}
\psi_{Dj}\bar{\psi}_{Dj}  \nonumber  \\
 \psi^h_{Si}\bar{\psi^h}_{Di}=
\sum_{j\neq i}  \psi_{Sj} \bar{\psi}_{Dj}
\end{eqnarray}

In the above analysis we have not considered the eigen-basis of the
detector. As we have considered the symmetrization aspects only, the
state $\psi_i$ occurs as a natural preferred vector and for the
other states $j\neq i$ we can consider any set orthogonal to
$\psi_i$.

\section{Quantum Images and the No-cloning Theorem}

The exchange interaction term due to (anti-)symmetrization
contains a product of the incoming microstate  $\psi_{Si}$, a
corresponding state $\psi_{iD}$ in the detector, which is the same
microstate for bosons, or a hole $\psi^h_{Di}$ in the case of
fermions, and also associated with such a pair is a conjugate
state  $\bar{\psi}_{Di}$ or $\bar{\psi^h}_{Di}$ for fermions. In
the case of the bosonic systems we shall call the latter conjugate
state an image of the original incoming state created by the
symmetrization process. We do not consider the symmetric identical
state $\psi_Di$ as the image, because the identical state
nominally in the detector is \emph{indistinguishable} from the
original incoming state and when there is an overlap of functions
they may represent the same physical entity. In the case of
fermionic systems the incoming state $\psi_{iS}$ and the
corresponding hole state $\psi^h_{Di}$ or its conjugate
$\bar{\psi^h}_{Di}$ are in general all nonidentical systems. Sine
the incoming state is definitely not $\psi^h_{Si}$, we can neglect
the second term in Eq. \ref{eq9}. Hence the effect of the
antisymmetrization effectively gives a simple product as for a
bosonic system:

\begin{equation} \label{eq11}
\Psi_{SD_{ferm}}  = \psi_{Si} \psi^h_{Di} \bar{\psi^h}_{Di}
\end{equation}

However, since the hole is more like a conjugate and the conjugate
of the hole is more like the original incoming  microsystem, we
can expect that both $\psi_{Si}$   and $\bar{\psi^h}_{Di}$
interact in a similar manner with $\psi^h_{Di}$.

There is no conflict with the no-cloning theorem \cite{ZU2} when
(anti-)symmetrization produces such quantum images, which, as we
have seen, are either extensions of the original functions, or are
conjugate states. Though there is a one-to-one correspondence with
the incoming state, the states in the detector simply extend the
original state by (anti-)symmetry or produce a state which is
conjugate to the original state, and is not producible by a
unitary operator assumed in the no-cloning theorem. In other
words, (anti-)symmetrization and the consequent exchange
interactions are not producible by the linear unitary operators
and the simple and elegant proof of the no-cloning theorem is
inappropriate for quantum images of the kind described above.

\section{Measurement and Eigenstates}

Quantum images, formed by invoking symmetrization properties of
the combined system, do not depend on the operator involved in the
measurement process associated with the detector. The quantity
measured is represented by a unitary operator in quantum
mechanics, and, if the microsystem is an eigenstate, it remains in
the same state even after measurement, but if it is a mixture of
eigenstates of the operator, then it is taken as a postulate of
quantum mechanics that the emerging state after measurement is one
of the eigenstates and the detector too carries off the
information of the final state to which it collapses. We have
shown recently \cite{SH1} how a first passage random walk model
reduces an arbitrary linear combination of eigenstates to one of
the component eigenstates with a probability proportional to the
square of the absolute magnitude of the coefficient of that
component. In that work we appealed to an electrostatic analogy
for the formation of the image in the detector which interacts
with the incoming microsystem in steps, both changing
simultaneously till an eigenstate is reached.

If the state $\psi_{Si}$ is expressed in terms of the eigenstates
in a simple two-state system

\begin{equation}\label{eq12}
\psi_{Si}=  a_i |\alpha \rangle_S  +  b_i |\beta \rangle_S
\end{equation}

then we get

\begin{equation} \label{eq13}
\psi_{Di} = a_i |\alpha \rangle_D  + b_i  |\beta \rangle_D
\end{equation}

and

\begin{equation} \label{eq14}
\bar{\psi}_{Di}= a^*_i |\bar{\alpha} \rangle_D  + b^*_i
|\bar{\beta} \rangle_D
\end{equation}

and similarly for the hole states in the case of the fermionic
systems.

This shows how the complex conjugate of the co-efficients occur in
a natural way in the image, which is not possible by cloning with
a unitary operator.

Here we also see that the conjugate can interact interchangeably
with the incoming state or its indistinguishable extension in the
detector and form virtual bound states

\begin{eqnarray}\label{eq15}
|SD\rangle_i \sim |a_i|^2 |\alpha \rangle_S |\bar{\alpha}\rangle_D
+|b_i|^2 |\beta
\rangle_S  |\bar{\beta}\rangle_D  \nonumber  \\
|DD\rangle_i \sim |a_i|^2 |\alpha \rangle_D |\bar{\alpha}\rangle_D
+|b_i|^2 |\beta \rangle_D  |\bar{\beta}\rangle_D
\end{eqnarray}

We have shown \cite{SH1} that first passage random walk in Hilbert
space with the modulus squared of the coefficients of the
eigenstate components of the microstate $(|a_{\alpha}|^2$,
$|b_{\beta}|^2, ...)$ as co-ordinates lead to unique final
eigenstates $|\alpha \rangle$, $|\beta \rangle$, ..., with
probabilities proportional to the corresponding initial
co-ordinates $(|a_{\alpha}|^2$, $|b_{\beta}|^2, ...)$, as
expected. The constant of proportionality is a measure of the
efficiency of the measuring device.

\section{Conclusions}

We have shown above that if the detector is a macroscopic system
and is initially neutral with respect to the measured quantity,
which we have expressed as the sum of microstates with all
different states, then symmetrization with the measured system for
bosonic systems or anti-symmetrization for fermionic systems
breaks the neutrality in a unique way which may be regarded as the
formation of a quantum image of the measured microsystem in the
detector. These images are conjugates of the incoming
microsystems, or hole-type states equivalent to conjugate states,
and since the process is not a linear unitary operation, the
no-cloning theorem does not pose a problem. That the interaction
between the incoming state and these images can be modeled by
first passage random walks to give the probabilities for different
eigenstates as final states of both the incoming state and the
detector's microstate component has been shown in \cite{SH1}. We
shall later examine the question of measurement of entangled
systems in spatially separated detectors.

\ack{

The author would like to thank  Professor R. Omn\`{e}s of
Universit\'{e} de Paris, Orsay, for useful feedback from the
earlier work.}

\end{document}